\documentclass[12pt]{iopart}
\usepackage{graphicx}
\begin{document}
\jl{3}
\title[CONQUEST]{Calculations on millions of atoms with DFT: Linear scaling shows its potential}
\author{D~R~Bowler$^{1,2,3}$ and T~Miyazaki$^{4}$}
\address{$^{1}$Thomas Young Centre, UCL, Gower St, London WC1E
  6BT, UK}
\address{$^{2}$London Centre for Nanotechnology, UCL, 17-19 Gordon St,
  London WC1H 0AH, UK}
\address{$^{3}$Department of Physics \& Astronomy, UCL, Gower St, London WC1E
  6BT, UK}
\address{$^{4}$National Institute for Materials Science, 1-2-1 Sengen,
  Tsukuba, Ibaraki 305-0047, JAPAN}
\eads{david.bowler@ucl.ac.uk}
\begin{abstract}
  An overview of the \textsc{Conquest} linear scaling density
  functional theory (DFT) code is given, focussing particularly on the
  scaling behaviour on modern high-performance computing (HPC)
  platforms.  We demonstrate that essentially perfect linear scaling
  and weak parallel scaling (with fixed atoms per processor core) can be
  achieved, and that DFT calculations on millions of atoms are now
  possible.
\end{abstract}
\submitted
\maketitle

\section{Introduction}
\label{sec:introduction}


Linear scaling approaches to atomistic calculations have their origin
in molecular dynamics codes with force fields: the idea that, by
calculating interactions for each atom only within a local part of
space, computational effort scales with a local volume leads to to
efficient increases in system size; this also leads to natural
parallelisation schemes.  It is well known that, for systems with a
gap or metals at finite temperature, electronic structure is local,
and falls off exponentially with distance---summed up in Kohn's
``nearsightedness'' principle\cite{kohn1996}.  This realisation led to
linear scaling tight binding methods in the 1980s.  Methods for $O(N)$
or linear-scaling DFT calculations\cite{goedecker1999} were first
proposed over 15 years
ago\cite{Yang:1991ei,Galli:1992vc,li1993,mauri1993,ordejon1993,nunes1994,Hern1995a},
but it is only in the last five years that practical calculations
using these methods have begun to appear.  
The developments which have enabled these calculations will be
surveyed in detail in Section~\ref{sec:methodology}.  In brief, the
algorithms used to find the ground state have converged on a few main
methods, while there has been more work on the local orbitals used to
represent the density matrix.  Local basis sets, and their efficient
implementation and minimisation, are key to performance in linear
scaling codes.

Part of the reason for the slow development of practical codes is that
parallelisation is extremely important.  If calculations on tens of
thousands or hundreds of thousands of atoms are to be performed, this
will require hundreds or thousands of processors (or cores in the case
of multi-core processors, as is becoming almost universal).  The
efficient implementation of linear scaling codes on parallel machines
has received attention
before\cite{Goringe1997a,Challacombe:2000vh,Bowler2001a,Brazdova:2008pz,Hine:2009ez};
in this paper, we will explore how far scaling can be extended
efficiently.  We find that there is every reason to believe that
linear scaling DFT will make extremely good use of the hundreds of
thousands of cores which are becoming available with petascale
computers\footnote{The Jaguar Cray machine installed at Oak Ridge
  National Laboratory in America is the first petaflop machine, and
  has 150,000 cores, while the next-generation supercomputer in Japan,
  which is scheduled for completion in 2011, will have a peak
  performance of over 10 petaflops and will require several hundred
  thousand cores.}.  In this article, we will consider the performance
of our linear scaling DFT code
\textsc{Conquest}\cite{Hern1995a,Goringe1997a,Hern1996a,Bowler1999a,Bowler2000b,Bowler2002b,Miyazaki2004a,Bowler2006a},
but there are other linear scaling DFT codes under development, for
instance Siesta\cite{soler2002}, OpenMX\cite{Ozaki2003a} and
ONETEP\cite{skylaris2005a}.  As will be described in the next section,
most linear scaling methods work by using a reformulation of DFT in
terms of the density matrix, and apply localisation constraints to
achieve good scaling with system size.

In the next section, we give an overview of the \textsc{Conquest}
methodology, covering the fundamental theory as well as details of the
implementation.  The results section forms the central part of the
paper, presenting scaling data both with respect to number of cores
and numbers of atoms.  We conclude with a brief look forward.

\section{Methodology}
\label{sec:methodology}

The ideas underlying \textsc{Conquest} have been presented
before\cite{Hern1995a,Goringe1997a,Hern1996a,Bowler1999a,Bowler2000b,Bowler2002b,Miyazaki2004a,Bowler2006a},
but we will give an overview here for convenience and to help explain
the implementation details given below; the interested reader is
referred to previous publications for full details.  As is common with
many linear-scaling codes, \textsc{Conquest} works directly with the
density matrix rather than wavefunctions, and writes it in a separable
form:

\begin{equation}
  \label{eq:1}
  \rho(\mathbf{r},\mathbf{r}^\prime) = \sum_{i\alpha, j\beta} \phi_{i\alpha}(\mathbf{r}) K_{i\alpha j\beta} \phi_{j\beta}(\mathbf{r}^\prime),
\end{equation}
where $\phi_{i\alpha}(\mathbf{r})$ is a strictly local function
centred on atom $i$ called a \emph{support function}; multiple support
functions on the same atom are notated with $\alpha$.  The support
functions are not orthogonal, and there is an associated overlap
matrix:
\begin{equation}
  \label{eq:2}
  S_{i\alpha j\beta} = \int \mathrm{d}\mathbf{r} \phi_{i\alpha}(\mathbf{r})\phi_{j\beta}(\mathbf{r}).
\end{equation}
The density matrix in the basis of support functions is written
$K_{i\alpha j\beta}$.  Locality is imposed in \textsc{Conquest} via a
spherical cutoff on the support functions $R_{\mathrm{cut}}$ and a
distance-based criterion on the elements of an auxiliary density
matrix from which $K$ is derived.

For a given set of support functions, the ground state is found by
varying the elements of $K$ to minimise the energy subject to various
conditions:

\begin{enumerate}
\item Self-consistency between the charge density and potential
\item Correct electron number, $N_e = 2\mathrm{Tr}[KS]$
\item Idempotency of the density matrix
\end{enumerate}

The first of these conditions is a standard problem within electronic
structure, and while not trivial, has been widely explored in other
contexts\cite{Bowler2006a,Johnson:1988fv,Kresse:1996fp}.  The second is relatively easy to
impose, and can be incorporated within the
minimisation\cite{Goringe1997a,Hern1996a}.  The final condition is
extremely hard to impose, and we instead use the ideas of
McWeeny\cite{mcweeny1960} to impose weak idempotency. By writing $K$
in terms of an auxiliary density matrix (ADM), $L$, we ensure that its
eigenvalues lie between 0 and 1 and converge towards these extrema as
the minimisation proceeds\cite{li1993,nunes1994,Bowler1999a,mcweeny1960,li1993}:

\begin{equation}
  \label{eq:3}
  K = 3LSL - 2LSLSL.
\end{equation}
This method for achieving idempotency is sometimes known as the ADM or LNV method.

Practically, the localisation on the density matrix is imposed on $L$,
so that $L_{ij} = 0, |\mathbf{R}_i - \mathbf{R}_j| > R_L$.  By using
sparse matrices, and carefully constructed sparse matrix
methods\cite{Bowler2001a}, the computational time and memory required
for minimisation of energy with respect to the elements of $K$ (and
ultimately $L$) scale linearly with the number of atoms in the
system.  

This is one area where \textsc{Conquest} differs from other linear
scaling DFT codes (though it is not the only linear scaling DFT code
to use the ADM method: the ONETEP code\cite{skylaris2005a}, for
instance, also uses it). The Orbital Minimisation Method
(OMM)\cite{ordejon1993,mauri1993,kim1995} is another variational
approach, though it is not commonly used
(it is implemented in the SIESTA code\cite{Soler2002a} and used by
Tsuchida\cite{Tsuchida:2007fx}).  Non-variational methods commonly
used include the divide-and-conquer\cite{Yang:1991ei} (D\&C) method
and the trace-correcting family of methods\cite{Niklasson:2002ef}).
These are the main methods used for the density matrix search.

The representation of the support functions or local orbitals is an
important problem within linear scaling electronic structure
techniques, and is another area where \textsc{Conquest} differs from
other codes.  The codes now
available\cite{Bowler2002b,Ozaki2003a,skylaris2005a,Soler2002a} are of
two types: those that use basis sets akin to plane waves (including
blips or B-splines\cite{Hern1997a}, finite element
approaches\cite{Tsuchida1998a,pask1999a,Pask2005a}, periodic sinc
functions\cite{Mostofi2002a} and wavelets\cite{Genovese:2008ya}),
which allow systematic basis-set convergence; and those that use
pseudo-atomic orbitals (PAOs) as basis
sets\cite{Ozaki2003a,Soler2002a,Sankey1989a,Kenny2000a,Torralba:2008wm,Blum:2009kw},
for which systematic convergence is usually significantly harder, but
which have smaller basis sizes. An important feature of our own
\textsc{Conquest}
code\cite{Bowler2002b,Miyazaki2004a,Bowler2006a} is
that both types of basis are implemented, and this means that rapid,
though semi-quantitative calculations can be performed for exploratory
purposes, but precise calculations are also possible.  The support
functions are written:
\begin{equation}
  \label{eq:4}
  \phi_{i\alpha}(\mathbf{r}) = \sum_s b_{i\alpha s} \chi_s(\mathbf{r})
\end{equation}
where $\chi_s(\mathbf{r})$ is a basis function centred on atom $i$.

The quantitative basis set uses blip functions, specifically
b-splines, on a cubic regular grid defined within the support region
for each atoms\cite{Hern1997a}, which can be related to a plane-wave
energy cutoff; it is, however, perfectly possible to use different
spacings for different atoms.  By increasing the support region radius
and the L matrix cutoff systematically, it is possible to achieve
plane-wave accuracy linear scaling
calculations\cite{Goringe1997a,Hern1997a,Skylaris:2007we}.  

The different computational operations in Conquest can be summarised as:

\begin{enumerate}
\item Matrix multiplication (e.g. $K_{ij} = L_{ik}S_{kl}L_{lj}$)
\item Integration on a grid (which is regular, and defined along the simulation cell lattice vectors)
\item Basis function operations: analytic integrals or basis-to-grid transformations
\item Fast fourier transforms (performed on the same grid as
  integration)
\item Communication of information between cores
\end{enumerate}

The parallelisation strategy in
\textsc{Conquest}\cite{Goringe1997a,Bowler2001a,Bowler2000b} relies on
the division of the computational cell into small groups of atoms
(partitions) and integration grid points (blocks); typically a
partition will contain $\sim$5--20 atoms, and a block will have size
of $3\times 3 \times 3$---$8\times 8\times 8$ grid points.  These are
assembled into groups (bundles of partitions and domains of blocks)
which should be both localised and overlapping for good
performance, and assigned to cores (for multi-core CPUs).  The
assembly of domains and bundles, and the assignment of these groups to
cores can strongly affect the efficiency of the code.  We have
implemented a default partitioning scheme based on Hilbert
curves\cite{Brazdova:2008pz} which allows calculations without
detailed optimisation of load balancing; examples of the effeciency of
parallel scaling with this scheme are given below in
Sec.~\ref{sec:results}.  Details of partitioning can be optimised
externally to Conquest, and this allows different approaches to be
taken.  There are some computational cells where the assignment of
domains and bundles to cores is obvious (for instance the cubic cells
used for scaling tests up to millions of atoms), and a simple script
will allow the optimal distribution to be created.  We also have an
optimising code which uses simulated annealing to load-balance the
system.

\textsc{Conquest} can operate at different levels of accuracy,
depending on the basis set chosen, and other factors.  If PAOs are
used, with only a minimal basis set and no self consistency, then we
have non-self-consistent \emph{ab initio} tight binding (NSC-AITB).
If self-consistency is introduced and the basis set expanded somewhat
then the code runs at the level of self-consistent AITB.  For full PAO
basis sets and blip functions with full basis optimisation we achieve
full DFT accuracy, and when cutoffs are taken to convergence we can
recover plane-wave accuracy.  \textsc{Conquest} has the PBE GGA
functional implemented as well as LDA, at all
levels\cite{Antonio-S.-Torralba:2009ye}.  Forces are calculated
exactly as derivatives of the
energy\cite{Miyazaki2004a,Antonio-S.-Torralba:2009ye} and are
implemented at all levels of accuracy, both for LDA and
GGA\cite{Antonio-S.-Torralba:2009ye}.

Many of the calculations in this paper operate at the NSC-AITB level,
as this uses the full functionality of the code and permits good
scaling tests.  This does not mean, however,
that this is how we anticipate using the code; indeed, we have performed
self-consistent calculations on cells up to 262,144 atoms (described 
below) with no decrease in the scaling.  Optimisation of support functions
scales in a similar manner.


\section{Results}
\label{sec:results}

The results in this section are intended to demonstrate the scaling
performance of the \textsc{Conquest} code.  We have recently used the
code for a series of calculations on the Ge(105)
surface\cite{Miyazaki:2007kr} and on the energetics of self-assembly
of Ge hut clusters on Si(001)\cite{Miyazaki:2008kx}, and we draw many
example systems from these studies.  Details of the systems are given
in the papers already published.  Linear scaling methods are also ideal for
application to ionic materials (which often have large band gaps) and
we have successfully performed exploratory self-consistent
calculations on MgO surfaces with defects.  We are also using
\textsc{Conquest} to perform calculations on other systems, such as
biomolecules, and we are actively pursuing this
area\cite{Antonio-S.-Torralba:2009ye,Otsuka:2008ri,Gillan:2007aq}

\begin{figure}[h]
  \centering
  \includegraphics[width=0.46\textwidth]{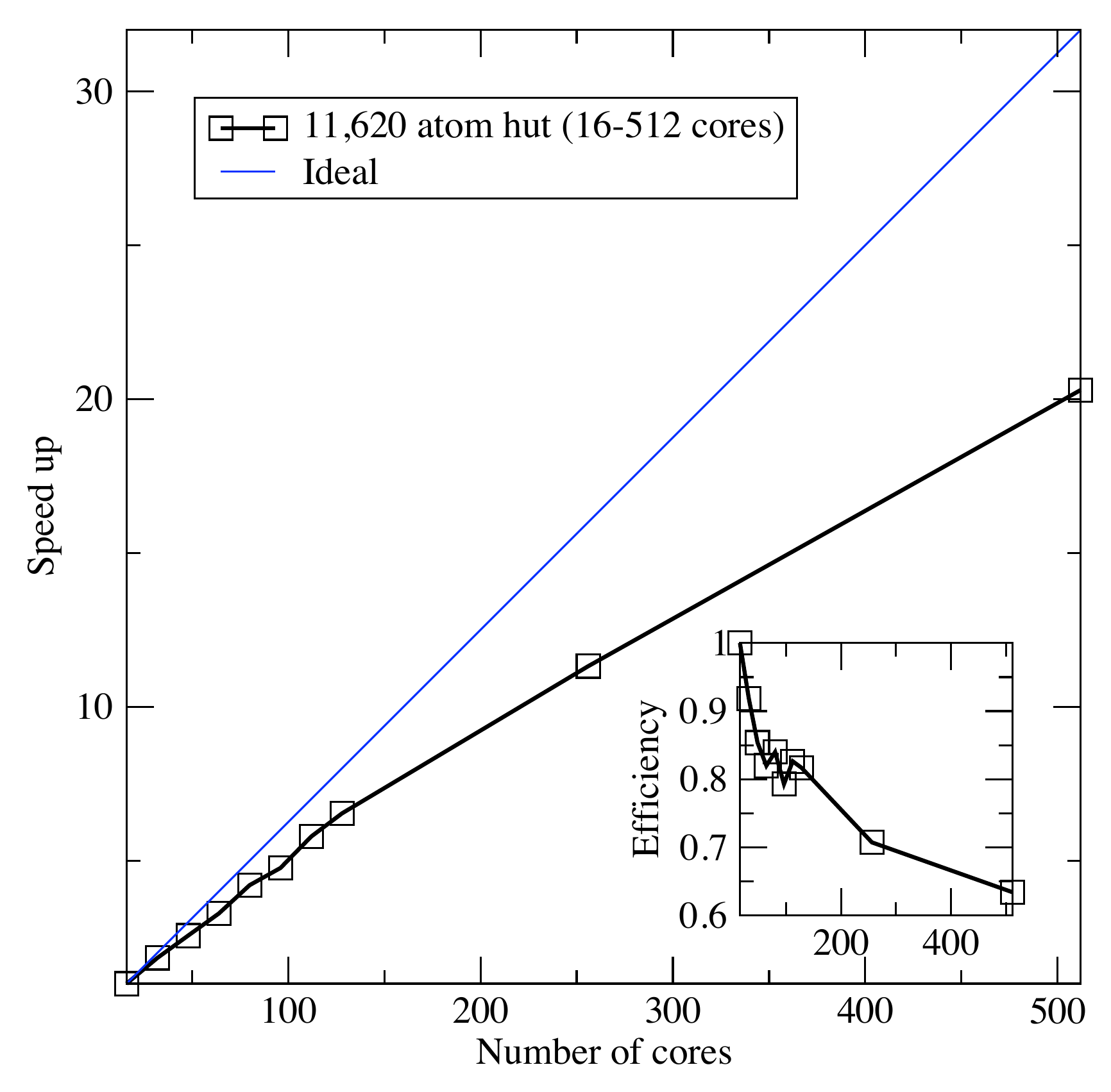}
  \includegraphics[width=0.45\textwidth]{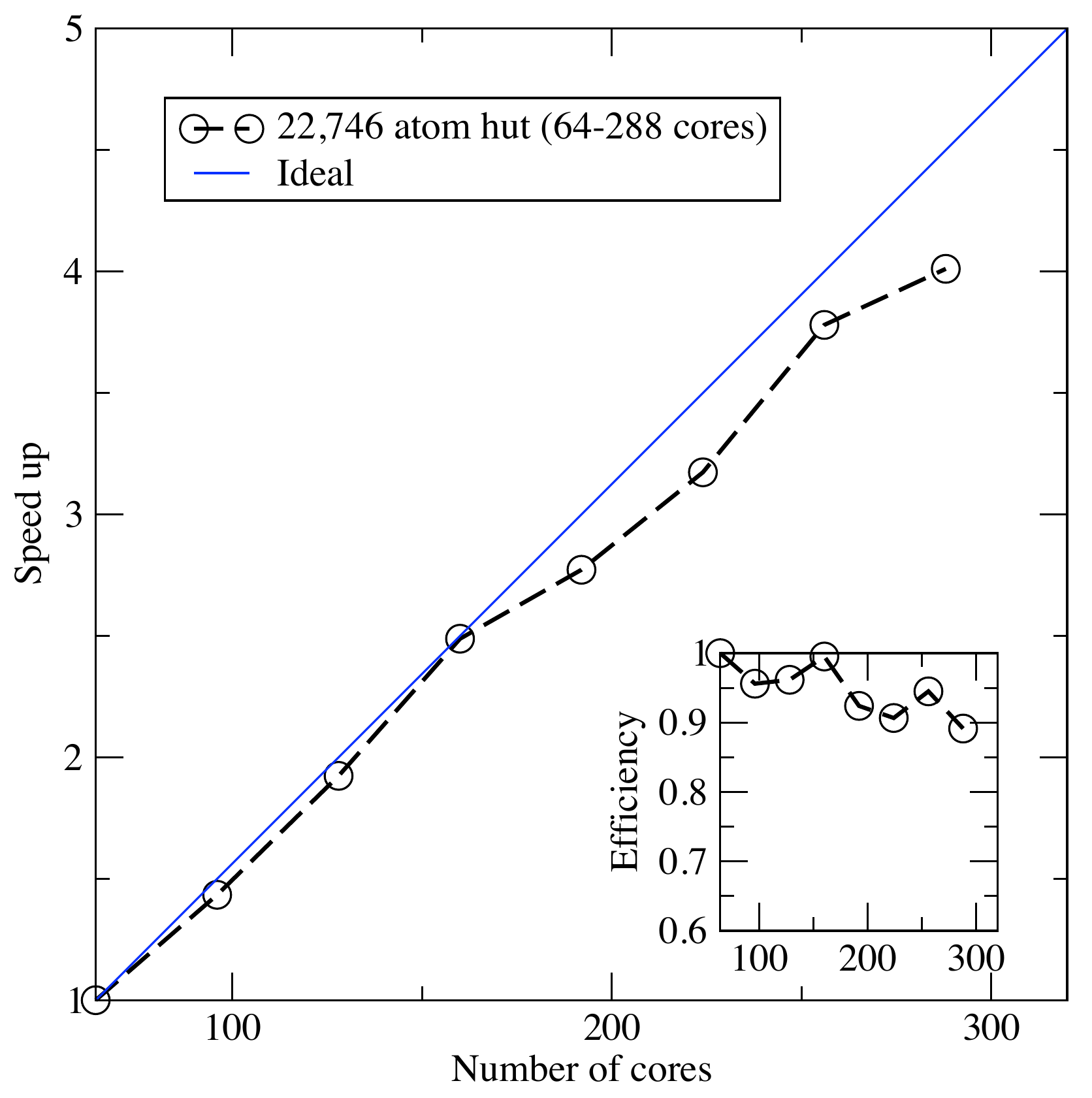}
  \caption{Scaling using automated Hilbert partitioning for two different hut clusters.  (a) Hut cluster with 11,620 atoms on 16---512 cores (increasing by 32 times); (b) Hut cluster with 22,746 atoms on 64---288 cores (increasing by 4.5 times).}
\label{fig:hutscalinghilbert}
\end{figure}

We start by considering the strong scaling performance on the current
UK HPC facility HECToR (a Cray XT4): that is increasing the number of
cores used in a calculation while keeping the system size fixed.  For
these tests, we have used two of the Ge hut cluster systems,
containing 11,620 and 22,746 atoms respectively.  The unit cells are
far from cubic, which presents a non-ideal situation for the default
partitioner which uses a 3D Hilbert curve and performs best for cells
close to cubic.\footnote{We are developing improvements to the
  partitioner to allow us to move away from this restriction, but
  these are still at a preliminary stage.}  The smaller system allows
somewhat better assignment of atoms to cores, and we have tested the
parallel scaling more extensively on this system.

Results are shown in Fig.~\ref{fig:hutscalinghilbert}.  In
Fig.~\ref{fig:hutscalinghilbert}(a) we show the speed up of the code
for the smaller hut cluster, as we increase the number of cores by a
factor of up to 32.  For an increase of a factor of up to 8, the
scaling is excellent with an efficiency (defined as speed-up measured
divided by increase in number of cores) of about 80\%.  For further
increases in numbers of cores we see smaller efficencies, but the
efficiency is still over 60\%.  In Fig.~\ref{fig:hutscalinghilbert}(b)
we show the scaling for the larger hut cluster, as the number of cores
is increased by a factor of up to 4.5.  This scaling is excellent, and
remains at over 90\%.

We can understand the strong scaling behaviour from the
parallelisation strategy.  The main computational load in
\textsc{Conquest} is the sparse matrix multiplication, which we have
optimised extensively\cite{Bowler2001a}.  The time required for
multiplies scales with the number of neighbours of each atom as well
as the number of atoms per core; in the hut cluster system shown
above, some atoms are near the surface of the system with fewer
neighbours, while others are in the bulk with more neighbours.  With
less than 20 atoms per core, it is rather hard to achieve good load
balancing.  Good load balancing also requires that the bundles of
atoms assigned to cores are compact, and this is difficult to achieve
with small numbers of atoms per core.  Also,
as the number of atoms per core decreases, communications overhead
will start to dominate.  This behaviour is clearly seen in
Fig.~\ref{fig:hutscalinghilbert}, where there are only $\sim$20
atoms/core at the largest number of cores.  The most efficient results
are seen for 40 or more atoms per core.

\begin{figure}[h]
  \centering
  \includegraphics[width=0.45\textwidth]{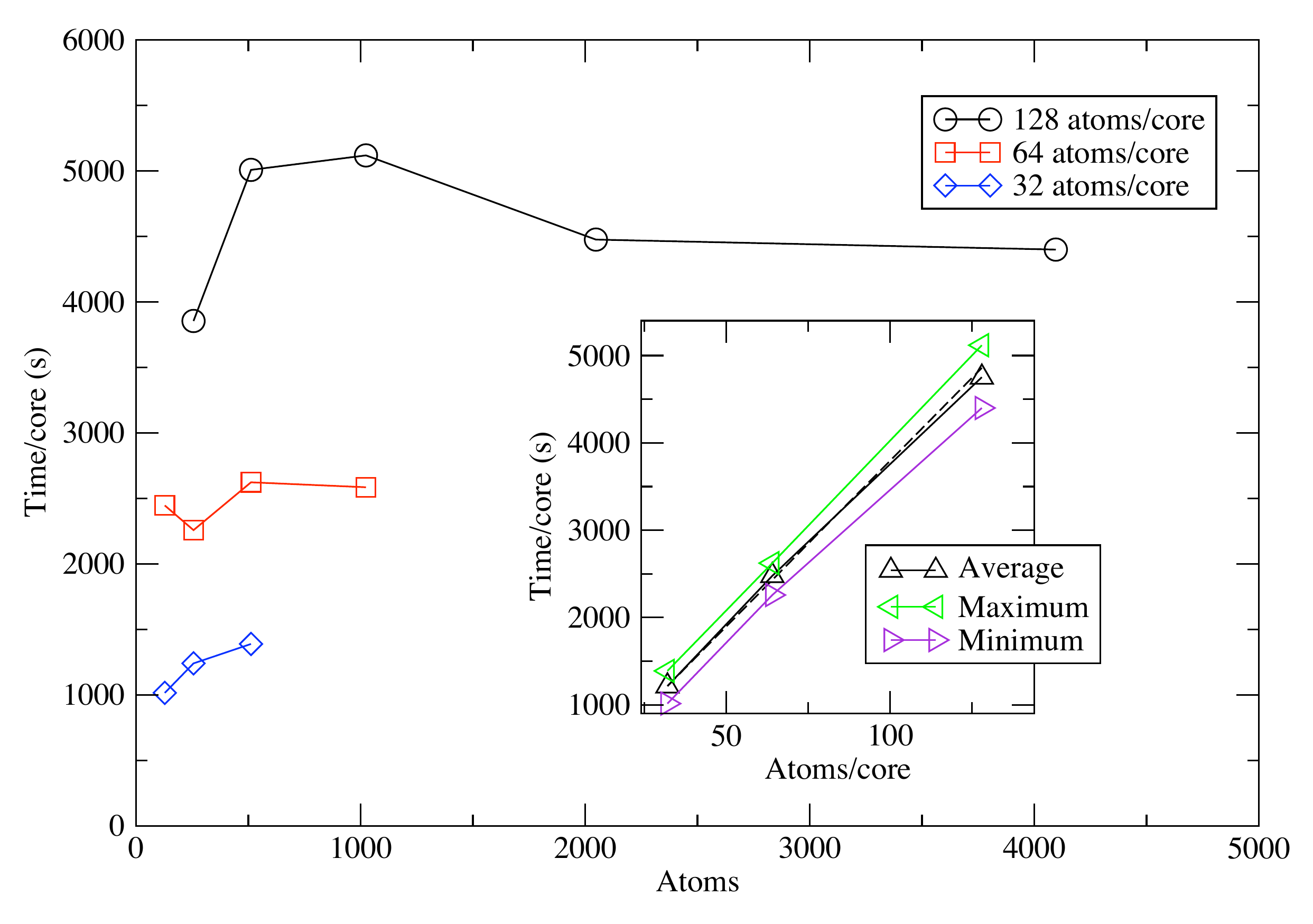}    
  \caption{Scaling on LCN cluster (dual core Opteron, Myrinet interconnect).  Main graph shows time per core for different atoms per core (reaching 4,096 atoms on 32 cores); inset shows the change of time/core with increasing numbers of atoms/core, with a linear increase for the average shown with a dashed line.}
  \label{fig:mithras}
\end{figure}

In practical calculations, we usually increase the number of cores
proportionately with the number of atoms in the system.  Therefore,
more realistic tests of the scaling are to fix the number of atoms per
core, and increase the number of cores at the same time as increasing
the number of atoms; this is known as weak scaling.  Results for this
mode of operation on a local HPC cluster (based on dual-processor,
dual-core Sun servers with Myrinet interconnects) are shown in
Fig.~\ref{fig:mithras}.  The system being tested is bulk silicon,
which while not scientifically interesting, is simple to prepare and
contains all the essential physics we wish to test.  The main graph
shows the time per core plotted against number of atoms in the system
for different numbers of atoms per core: 32 atoms/core, 64 atoms/core
and 128 atoms/core.  A number of important points come out of this
plot: first, the time per core is effectively constant for the systems
considered; second, communication becomes unimportant for 64
atoms/core or more (as seen in Fig.~\ref{fig:hutscalinghilbert} as
well); third, as shown in the inset, the linear scaling performance of
the code is excellent, lying on the ideal linear curve.

\begin{figure}[h]
  \centering
  \includegraphics[width=0.45\textwidth]{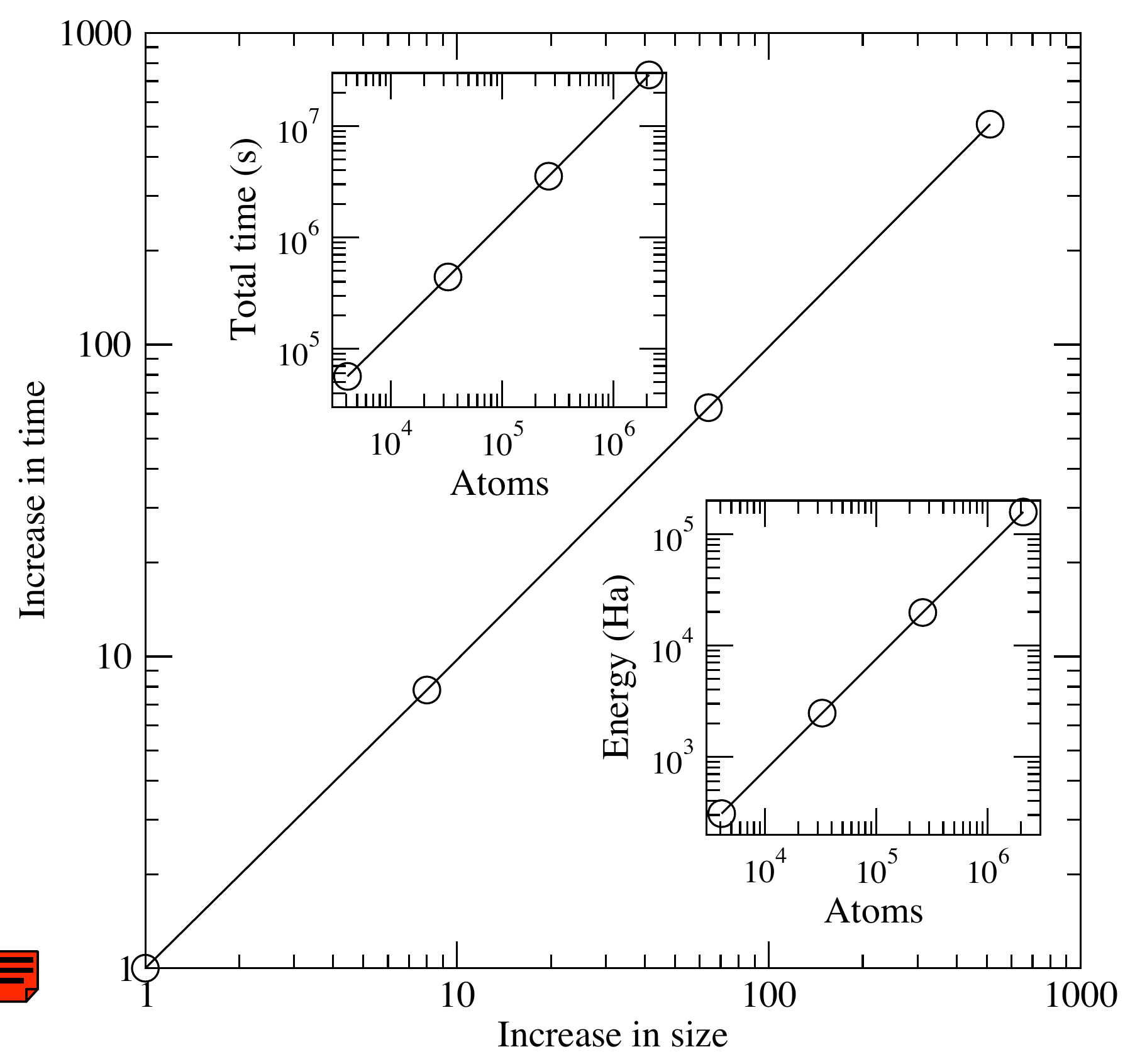}  
  \caption{Linear and parallel scaling for bulk silicon on 512---4096 cores.  The insets show total time and total energy (made positive to enable log plot) while main graph shows increase in time with system size.  Exact data values are given in Table~\protect\ref{tab:timemassive}.}
  \label{fig:massive}
\end{figure}

Finally, we are concerned to show that this good scaling behaviour
persists to extremely large systems, so we have taken a system which
can be easily partitioned and scaled and scaled to over 4,000 cores
and over 2,000,000 atoms.  We show increase in total time
(i.e. time/core summed over cores) vs increase in system size, as well
as total time and total energy plotted against number of atoms in
Fig.~\ref{fig:massive}.  These were run on HECToR, using between 8 and
4,096 cores with 512 atoms/core, giving 2,097,152 atoms as the largest
cell considered.  Details of times, energies and numbers of cores are
given in Table~\ref{tab:timemassive}.\footnote{The grid spacing used was a little coarser than we would normally choose, to reduce memory requirements; however, we note that this will not affect the convergence or scaling, and have tested the smaller systems with finer grids to ensure that there is no effect from this.}

\begin{table}[h]
  \centering
  \begin{tabular}{rrrr}
Atoms & Time/core (s) & Total energy (Ha) & Cores\\
      4,096 &    7068.878&      -308.268785          &       8    \\
    32,768  &   6893.759&    -2,466.150282        &    64    \\
  262,144   &  6931.418&  -19,729.202254      &   512    \\
2,097,152    & 7032.496 & -157,833.618033    & 4096\\
  \end{tabular}
  \caption{Times and energies for Conquest runs with 512 atoms/core.  The energy per atom takes a constant value of 0.075261 Ha.}
  \label{tab:timemassive}
\end{table}

The most important result from this calculation is that DFT
calculations on millions of atoms are now possible.  We see that the
time per core does not increase with system size, and that the energy
per atom is constant.  There are parts of the \textsc{Conquest} code
which are not strictly linear scaling: we use an Ewald sum for
electrostatic interactions (which can be easily replaced with a scheme
such as the neutral atom potential\cite{Soler2002a,Sankey1989a}) which
scales as $\mathcal{O}(N^{3/2})$ and fast fourier transforms which
scale as $N \mathrm{log}(N)$, but even for the 2,097,152 atom unit
cell these are negligible (approximately 3s for all FFT-related work
and 50s for Ewald sum).  We note that orbital-free DFT calculations
have recently been performed on a supercell of 1,012,500 atoms of bulk
Al\cite{Hung:2009yl}.

We have also performed self-consistent O(N) calculations on this
system (actually for the first three cells), and find that they
require four to five times as long, with our current implementation
(in this case, the variational nature of the minimisation means that,
as self-consistency is approached, less time is spent finding the
density matrix); the scaling of the code when performing
self-consistent calculations, is identical to non-self-consistent
calculations.  The main challenge now is to improve the efficiency of
the code, and reduce the number of atoms per core which can be run
without communciations becoming a heavy burden.  We will focus on
three main areas: first, efficient re-use of variational data such as
the L matrix to reduce the time to the ground state; second,
robustness and stability of the calculations; and finally, more
efficient automatic partitioning\cite{Brazdova:2008pz}.  This will allow us to
consider real scientific problems which require tens or hundreds of
thousands of atoms, and to perform molecular dynamics simulations
using petascale computer platforms.  


\section{Conclusions}
\label{sec:conclusions}

Linear scaling approaches to DFT have been under development for about
fifteen years, and are now starting to show their promise in real
calculations, and in their applicability to petascale computers.  This
special issue of Journal of Physics: Condensed Matter is in honour of
Professor Mike Gillan's 65th birthday, and it is appropriate to
celebrate the considerable contribution which he has made to the
development of linear scaling DFT, both through the theory and
implementation\cite{Hern1995a,Goringe1997a,Bowler2001a,Hern1996a,Bowler1999a,Bowler2000b,Bowler2002b,Miyazaki2004a,Bowler2006a,Hern1997a,Miyazaki:2007kr,Miyazaki:2008kx,Otsuka:2008ri,Bowler1998a,Bowler2002a,Torralba:2008wm}.
The results in this paper show that linear scaling DFT is realising
its potential, and that Mike Gillan's contributions have underpinned
all that has gone on in the field.

It has been noted\cite{Bowler:2008jh} that applications of linear
scaling methods to real problems are starting to emerge; the 
challenge now is to make linear scaling methods sufficiently robust and
efficient that they can be used as routinely as standard DFT methods, 
and to find applications which demonstate their power.
Examples of applications of these methods include work on DNA with
Siesta\cite{Pablo:2000ph} and \textsc{Conquest}\cite{Otsuka:2008ri},
biomolecules with ONETEP\cite{Heady:2006fm} and
\textsc{Conquest}\cite{Antonio-S.-Torralba:2009ye,Gillan:2007aq} and
our work on Ge on Si(001) with \textsc{Conquest}, extending to over
20,000 atoms\cite{Miyazaki:2007kr,Miyazaki:2008kx}.  Among other
applications, we intend to extend the Ge work to the transition from
hut clusters to domes, as well as applying \textsc{Conquest} to
biomolecules\cite{Gillan:2007aq}.  The code will also be released
under a GPL licence in the near future.



\ack

DRB is supported by the Royal Society.  This work is partly supported
by Grant-in-Aid for Scientific Research from the MEXT and JSPS, Japan.
Computational work was performed on the LCN cluster and HECToR under
the UKCP consortium.  Chris Goringe, Eduardo Hern\'andez, Ian Bush,
Rathin Choudhury, Veronika Br\'azdov\'a, Milica Todorovi\'c, Takao
Otsuka and Antonio Torralba are all acknowledged for valuable
contributions to the \textsc{Conquest} project.  We also thank Mike
Gillan for the on-going pleasure of working with him on the
\textsc{Conquest} project, as well as many years of supervision,
advice and support.

\section*{References}

\bibliographystyle{jpcm}
\bibliography{Conquest,GeneralON,PAOs}

\end{document}